\documentclass[journal,10pt,onecolumn,draftclsnofoot]{ieeetran}
\usepackage[T1]{fontenc}% optional T1 font encoding
\usepackage{url}
\usepackage{graphicx}
\ifCLASSINFOpdf
\else
\fi
\usepackage{cite}
\usepackage{amsmath}
\usepackage{color}

\usepackage[cmintegrals]{newtxmath}
\usepackage[noend]{algpseudocode}
\usepackage{subfigure}

\usepackage{algorithmicx,algorithm}
\usepackage[justification=centering]{caption}

\begin{document}
	% paper title
	% Titles are generally capitalized except for words such as a, an, and, as,
	% at, but, by, for, in, nor, of, on, or, the, to and up, which are usually
	% not capitalized unless they are the first or last word of the title.
	% Linebreaks \\ can be used within to get better formatting as desired.
	% Do not put math or special symbols in the title.
	\title{Full-Dimensional Rate Enhancement for\\ UAV-Enabled Communications via Intelligent Omni-Surface}
	%
	% author names and IEEE memberships
	% note positions of commas and nonbreaking spaces ( ~ ) LaTeX will not break
	% a structure at a ~ so this keeps an author's name from being broken across
	% two lines.
	% use \thanks{} to gain access to the first footnote area
	% a separate \thanks must be used for each paragraph as LaTeX2e's \thanks
	% was not built to handle multiple paragraphs
	%
	\author{	Yifan Liu,~\IEEEmembership{Student Member, IEEE,}
		Bin Duo,~\IEEEmembership{Member, IEEE,}
		Qingqing Wu,~\IEEEmembership{Senior Member, IEEE,}\\
		Xiaojun Yuan,~\IEEEmembership{Senior Member, IEEE,}
		and Yonghui Li,~\IEEEmembership{Fellow, IEEE}
		% <-this % stops a space
	
		\thanks{Yifan Liu and Bin Duo are with Chengdu University of Technology, Chengdu, China (e-mail:liuyifan@stu.cdut.edu.cn; duobin@cdut.edu.cn). Qingqing Wu is with the State Key Laboratory of Internet of Things
			for Smart City, University of Macau, Macau, China (email:
			qingqingwu@um.edu.mo). Xiaojun Yuan is with the National Laboratory of Science and Technology on
			Communications, University of Electronic Science and Technology of China,
			Chengdu, China (e-mail: xjyuan@uestc.edu.cn). Yonghui Li is with the School
			of Electrical and Information Engineering, University of Sydney, NSW, Australia (yonghui.li@sydney.edu.au). } }% <-this % stops a space
	\maketitle

	\begin{abstract}
		This paper investigates the achievable rate maximization problem of a downlink  unmanned aerial vehicle (UAV)-enabled communication system aided by an intelligent omni-surface (IOS). Different from  the state-of-the-art reconfigurable intelligent surface (RIS) that only reflects  incident signals, the IOS can simultaneously reflect and transmit the signals, thereby providing full-dimensional rate enhancement. We formulate this rate maximization problem  by jointly optimizing the IOS's phase shift and the UAV trajectory. Although it is difficult to solve it optimally due to its non-convexity, we propose an efficient algorithm to obtain a high-quality suboptimal solution.  Simulation results show that the IOS-assisted UAV communications can achieve significant improvement in achievable rates compared to other benchmark schemes.
	\end{abstract}
	
	% Note that keywords are not normally used for peerreview papers.

	\newpage

	% For peer review papers, you can put extra information on the cover
	% page as needed:
	% \ifCLASSOPTIONpeerreview
	% \begin{center} \bfseries EDICS Category: 3-BBND \end{center}
	% \fi
	%
	% For peerreview papers, this IEEEtran command inserts a page break and
	% creates the second title. It will be ignored for other modes.
	\IEEEpeerreviewmaketitle

	\section{Introduction}
	% The very first letter is a 2 line initial drop letter followed
	% by the rest of the first word in caps.
	% 
	% form to use if the first word consists of a single letter:
	% \IEEEPARstart{A}{demo} file is ....
	% 
	% form to use if you need the single drop letter followed by
	% normal text (unknown if ever used by the IEEE):
	% \IEEEPARstart{A}{}demo file is ....
	% 
	% Some journals put the first two words in caps:
	% \IEEEPARstart{T}{his demo} file is ....
	% 
	% Here we have the typical use of a "T" for an initial drop letter
	% and "HIS" in caps to complete the first word.
%	\IEEEPARstart{}{}
    With the commercialization of the fifth-generation (5G) wireless networks, the sixth generation (6G) wireless communication technology has recently attracted increasing attention by industry and academia \cite{9003618}. \textcolor{blue}{The reconfigurable intelligent surface (RIS), as an
   effective solution to overcome non-line-of-sight (NLoS) transmission
    and  coverage blind spots, can significantly improve the
    spectral efficiency, energy efficiency, security, and reliability of
    communication systems \cite{9326394}.} Thus, the  RIS has become a promising technology for the future 6G networks. In general, the RIS  is an artificial surface made of electromagnetic material and  consists of a large number of square metal patches, each of which can be digitally controlled to induce different reflection amplitudes and phases on incident signals \cite{777}. By optimizing the RIS's phase shifts, the signals from different transmission paths can be aligned at the desired receiver to boost  achievable rates \cite{09}.
	
	On the other hand, due to line-of-sight (LoS) transmission and flexible mobility, unmanned aerial vehicles (UAVs) have gained widespread attention and have been applied in many applications \cite{9055054,8918497,DB}. Thanks to the low-profile and lightweight of RISs, they can be installed at proper locations to  reconfigure propagation environments of air-to-ground channels, thus significantly enhancing communication quality. As shown in \cite{8959174} and \cite{glnz}, the average achievable (secrecy) rates of RIS-assisted UAV communications can be significantly increased by jointly optimizing the UAV trajectory and the RIS's phase shifts. To provide  better quality-of-service (QoS) for ground nodes that locate far apart,  a UAV relaying system with the aid of a RIS was proposed in \cite{tly}, where the relaying UAV forwards the received signals reflected from the RIS  via intelligent phase alignment. The advantages of employing multiple RISs for enhancing the received power of the UAV-enabled communications were further investigated in \cite{cxk}. Generally, the RIS can be placed on the outer surface of a building. In this case, the RIS can only reflect incident signals towards ground nodes. When the ground nodes are located indoors or in the back side of the RIS, their achievable rates may not benefit from the deployment of the RIS.
	
	To address the above issue, the intelligent omni-surface (IOS) has been proposed as an upgrade of the RIS to realize the dual functionality of signal reflection and transmission \cite{networks}. Similar to the RIS, the IOS is made of multiple passive scattering elements and programmable PIN diodes, which can be appropriately designed and configured to customize the propagation environment \cite{9200683}. Specifically, to realize full coverage of  ground users, the IOS-assisted  system was considered in \cite{9200683}, where the spectral efficiency is increased significantly by optimizing the phase shifts of the IOS. In \cite{123}, multiple indoor users obtain omni-directionally services from a small base station (SBS) with the aid of an IOS. By jointly optimizing the IOS analog beamforming and the SBS digital beamforming,  the received power of multiple users is enhanced. Due to the advantages of the IOS for both reflection and transmission, it is essential to investigate whether the application of the IOS to the UAV communication system could provide higher achievable rates for ground nodes in all directions.
	
	Motivated by the above, this paper considers a typical application scenario for UAV-enabled communications in 6G networks, where the UAV as  an aerial base station (BS) provides communication services to a ground node. To achieve omini-directional rate enhancement, an IOS is installed to intelligently reflect/transmit incident signals from the UAV. We aim to maximize the average achievable rate by jointly optimizing the UAV trajectory and the phase shift design of the IOS.  To resolve the non-convexity of the formulated problem, we develop an efficient algorithm to obtain a suboptimal solution.  Simulation results show that significantly higher achievable rates can be obtained by replacing the RIS with the IOS in the UAV-enabled communication system.  \textcolor{blue}{Note that we are the first to introduce the IOS to the UAV communication system for providing the full-dimensional rate enhancement. Furthermore, due to the complex physical characteristics and the radiation pattern of the IOS, both the formulated objective function and our proposed algorithm are different from those in the conventional the RIS-aided UAV communication system [8]-[11].}

	\section{System Model and Problem Formulation}
In this paper, we consider a typical application scenario for UAV-enabled communications, where a UAV flies on a mission to provide communication services to a ground node\footnote{\textcolor{blue}{Due to the space limitation, we consider the single ground node scenario. However, it can be readily extended to the case with multiple ground nodes by optimizing the communication scheduling for multiple ground nodes, while without changing the phase shift design of the IOS and the trajectory design of the UAV.}} (G). Owing to intelligent reflection as well as transmission of arrived signals, an IOS is deployed to  enhance G's achievable rate in all directions. We characterize the position of the UAV, the IOS, and G via the three-dimensional Cartesian coordinate system. As such, it is assumed that the UAV flies at a fixed altitude $z_U$ to communicate with G, whose location is denoted by $\mathbf{w}_G=[x_G, y_G]$, over a duration $T$. For convenience, $T$ is divided into $N$ time slots that are equal in length, i.e. $T=N \delta_{t}$, where $\delta_{t}$ is the  length of each time slot. Thus, the horizontal trajectory of the UAV can be approximated by the discrete way-points $\mathbf{q}[n]=[x[n], y[n]]$, $n \in \mathcal{N}=\{1, \cdots, N\}$. 
%	\begin{figure}[t]
%	\centering
%	\includegraphics[width=60mm]{zhutou.pdf}\\
%	\caption{An IOS-assisted UAV communication system. }
%	\label{fig:env}
%	\vspace{-3mm}
%\end{figure} 	
	%\begin{equation}
	%	\|\mathbf{q}[n+1]-\mathbf{q}[n]\|^{2} \leq D^{2}, n=1, \cdots, N-1,  \tag{2}
	%\end{equation}
	%\begin{equation}
	%\mathbf{q}[N]=\mathbf{q}_{F}, \mathbf{q}[1]=\mathbf{q}_{0}, \tag{3}
	%\end{equation}
	% where $\mathbf{q}_{0}$ and $\mathbf{q}_{F}$ denote the initial and final horizontal positions of the UAV respectively, $D = v_{max}\delta_{t}$ is the maximum distance the UAV can move horizontally in a time slot, and $v_{max}$ is the maximum speed of the UAV.
	We assume that both the UAV and G are equipped with a single omni-directional antenna. For the IOS with $M$ elements  \cite{9200683}, we denote by $\mathbf{w}_m=[x_m, y_m]$,  $m \in \mathcal{M}=\{1, \cdots, M\}$ and $z_m$ the horizontal and vertical coordinates of the $m$th element of the IOS, respectively. Thus, the normalized power radiation patterns of the arrival signal and the departure signal of the IOS can be expressed as  \cite{9200683}
	\begin{align*}
		&K_{m}^{A}[n]=\left|\cos ^{3} \theta_{m}^{A}[n]\right|=\left|\left(\frac{x[n]-x_m}{d_{U, m}[n]}\right)^{3}\right|,\theta_{m}^A[n] \in (0,2\pi),\quad (1)&\nonumber
	\end{align*}
	\begin{align*}
		&K^{D}_m=\left\{\begin{array}{l}
			\left|\cos ^{3} \theta_{m}^{D}\right|=\left|\left(\frac{x_{G}-x_{m}}{d_{m,G}}\right)^{3}\right|,\theta_{m}^D \in (0,\frac{\pi}{2}),\\
			\epsilon\left|\cos ^{3}\left(\pi-\theta_{m}^{D}\right)\right|=\epsilon\left|\left(\frac{x_{G}-x_{m}}{d_{m, G}}\right)^{3}\right|, \theta_{m}^D \in (\frac{\pi}{2},2\pi),\quad(2)
		\end{array}\right.&\nonumber
	\end{align*}
 	where $\theta_{m}^{A}[n]$ is the angle of arrival (AoA) of the signal from the UAV to $m$th element of the IOS in the $n$th time slot, $\theta^{D}_m  $ is the angle of departure (AoD) of the signal from the $m$th element to G, $\epsilon$ is a constant that is determined by the hardware structure of the IOS \cite{2013}, $d_{U,m}[n]=\sqrt{\left\|\mathbf{q}[n]-\mathbf{w}_{m}\right\|^{2}+(z_U-z_m)^{2}}$  and $d_{m,G}=\sqrt{\left\|\mathbf{w}_G-\mathbf{w}_{m}\right\|^{2}+z_m^{2}}$ are the distances from the $m$th element of the IOS to the
	UAV and to G, respectively. Thus, the reflective or transmissive power gain from the $m$th element of the IOS to G is given by  \cite{9206044}
	\begin{align*}
		g_{m}[n]=\sqrt{G_{m} K^{A}_m[n] K_{m}^{D} \delta_{x} \delta_{y}\left|\gamma_{m}\right|^{2}} \exp \left(-j \psi_{m}[n]\right), \tag{3}
	\end{align*}
	where $G_m$ is the antenna power gain of the $m$th reconfigurable unit, $\delta_y$ and $\delta_z$ are the size of each element along $Y$ and $Z$ aixs, respectively,  $|\gamma_{m}|^2$ is the power ratio between the power of the signal
reflected/transmitted by the IOS and the power of the incident signal, and $\psi_{m}[n]$ denotes the phase shift of the $m$th element in time slot $n$.
%	the direction of the signal that is emitted by the UAV and that impinges upon the $m$th
%	reconfigurable element of the IOS in the $n$ time slot is denoted by $\xi^{A}_m[n]=\left(\theta^{A}_m[n], \phi^{A}_m[n]\right)$, where $\phi^{A}_m[n]$ represents the Angle of arrival of azimuth in the $n$ time slot. Futhermore, the direction of
%	the signal that is re-emitted by the $m$th reconfigurable element of the IOS towards the G
%	is denoted by $\xi_{m}^{D}=\left(\theta_{m}^{D}, \phi_{m}^{D}\right)$, where $\phi_{m}^{D}$ represents the Angle of departure of azimuth. 
	 
	 In this system, the UAV communicates with G via two links: the direct path from the UAV to G and the reflective-transmissive path from the UAV to G via the IOS. It is assumed that the channel coefficients from the UAV to G follow a practical Rician channel model, i.e.,
	\begin{equation}
		h_{D}[n]=\sqrt{\frac{\kappa}{1+\kappa}} h_{D}^{\operatorname{LoS}}[n]+\sqrt{\frac{1}{1+\kappa}} h_{D}^{N \operatorname{LoS}}[n], \tag{4}
	\end{equation}
	where $\kappa$ is the Rician factor. Furthermore, $h_{D}^{\operatorname{LoS}}[n]=\sqrt{G^{tx} G^{rx} d_{U, G}^{-\alpha}[n]} \exp \left(-j \frac{2\pi}{\lambda} d_{U,G}[n]\right)$ is the deterministic LoS
	component, where $G^{tx}$ is the transmission antenna gain of the UAV antenna,  $G^{rx}$ is the receiving antenna gain of G, $d_{U,G}[n]=\sqrt{\left\|\mathbf{q}[n]-\mathbf{w}_{G}\right\|^{2}+z_U^{2}}$ is the distance from the UAV to G, and $\alpha$ is the path-loss exponent. The NLoS component is defined as $h_{D}^{N L o S}[n]=\sqrt{G^{tx} G^{rx} d_{U, G}^{-\alpha}[n]} h_{S S}$, where $h_{S S} \sim \mathcal{C} \mathcal{N}(0,1)$ is the small-scale fading component modeled by a circularly symmetric complex
	Gaussian (CSCG) random variable. Similarly, the channel coefficients from the UAV to G via the $m$th IOS element can also be formulated as a Rician channel, which is given by
	\begin{equation}
		h_{m}[n]=\sqrt{\frac{\kappa}{1+\kappa}} h_{m}^{\operatorname{LoS}}[n]+\sqrt{\frac{1}{1+\kappa}} h_{m}^{ \operatorname{NLoS}}[n]. \tag{5}
	\end{equation}
The LoS component of $h_m[n]$ can be expressed as
	\begin{align*}
		h_{m}^{\operatorname{LoS}}[n]&=\frac{\lambda \sqrt{G^{t x} K_{m}^{A}[n] G^{r x} K_{m}^{D}} \exp \left(\frac{-j 2 \pi\left(d_{U, m}[n]+d_{m, G}\right)}{\lambda}\right)}{(4 \pi)^{\frac{3}{2}} d_{U A V, m}[n] d_{m, M U}} \times g_{m}[n]\\
		&=\frac{\lambda K_{m}^{A}[n] K_{m}^{D} \sqrt{G^mG^{t x}  G^{r x} \delta_{z} \delta_{y}\left|\gamma_{m}\right|^{2}}}{(4 \pi)^{\frac{3}{2}} d_{U, m}[n] d_{m, G}}\times\exp \left(\frac{-j 2 \pi\left(d_{U, m}[n]+d_{m, G}+\psi_{m}[n]\right)}{\lambda}\right), \tag{6}
	\end{align*}
	where  $\lambda$ is the carrier wavelength. The NLoS component of $h_m[n]$ can be represented as 
	\begin{align*}
		h_{m}^{N L o S}[n]=\frac{\lambda K_{m}^{A}[n] K_{m}^{D} \sqrt{G^mG^{t x}  G^{r x} \delta_{z} \delta_{y}\left|\gamma_{m}\right|^{2}}}{(4 \pi)^{\frac{3}{2}} d_{U, m}[n] d_{m, G}}  h_{S S}. \tag{7}
	\end{align*}
	Therefore, the channel coefficient from  the UAV to G via the IOS in the $n$th time slot can be given by 
	\begin{equation}
		h[n]=\sum_{m=1}^{M} h_{m}[n]+h_{D}[n]. \tag{8}
	\end{equation}

	It is assumed that the UAV transmits with its maximum power denoted by $P$. Then, the average achievable rate in bits/second/Hertz
	(bps/Hz) at G in the $n$th time slot is given by
	\begin{equation}
		\bar{R}=\dfrac{1}{N}\sum_{n=1}^{N}\log _{2}\left(1+ \eta |h[n]|^{2}\right), \tag{9}
	\end{equation}
	where $\eta=\dfrac{P}{\sigma^2}$, and $\sigma^2$ is the additive white Gaussian noise power
	at the corresponding receiver.
	
	Our goal is to maximize the average achievable rate by jointly optimizing the horizontal UAV trajectory $\mathbf{Q} \triangleq\{\mathbf{q}[n], n \in \mathcal{N}\}$ and the IOS's phase shift  $\mathbf{\Psi} \triangleq\{\mathbf{\psi}_m[n], n \in \mathcal{N}, m \in \mathcal{M}\}$ over the entire $N$ time slots. Therefore, the optimization problem can be expressed as
	\begin{align*}
		\max _{\mathbf{Q}, \mathbf{\Psi}}\quad &\bar{R} \tag{10a}\\
		\text { s.t. }&\|\mathbf{q}[n]-\mathbf{q}[n-1]\|^{2} \leq D^{2},\forall n,  \tag{10b} \\
		&\mathbf{q}[N]=\mathbf{q}_{F}, \mathbf{q}[1]=\mathbf{q}_{0},  \tag{10c}\\
		&0\le\psi_{m}[n]\le2\pi,\forall n,m,  \tag{10d}
	\end{align*}
where $\mathbf{q}_0$ and $\mathbf{q}_F$ denote the initial and final horizontal positions
of the UAV, respectively, $D = v_{max}\delta_{t}$ is the maximum 
 distance that the UAV can move horizontally within a time slot, and $v_{max}$ is the maximum  airspeed of the UAV.
	Problem (10) is difficult to solve because it is a complicated non-linear fractional programming problem with an objective function that is not jointly concave with respect to its coupled optimization variables. In the following section, we will propose an alternating optimization algorithm to solve problem (10).
	\section{Proposed Algorithm}
	In this section, we propose an alternating optimization algorithm to solve problem (10) by alternately optimizing the IOS's phase shift $\mathbf{\Psi}$ and the UAV trajectory $\mathbf{Q}$ until the convergence of the algorithm.
	\subsection{IOS Phase Shift Design}
	With any feasible UAV trajectory $\mathbf{Q}$, the optimization problem of phase-shift $\Psi$ can be expressed as
	\begin{align*}
		&\max _{\mathbf{\Psi}} |h[n]|^{2} \tag{11}\\
		&\quad\text{s.t. (10d).} 
	\end{align*}
\textcolor{blue}{Optimized $\Psi$ allows the signals from different paths to be combined coherently at $\mathrm{G}$, thereby maximizing its average achievable rate. Since $h_{m}^{\mathrm{NLoS}}[n]$ and $h_{D}^{\mathrm{NLoS}}[n]$ are both constants and both nonnegative, maximizing $|h[n]|^{2}$ is to maximize $\left|\sum_{m=1}^{M} h_{m}^{\mathrm{LoS}}[n]+h_{D}^{\mathrm{LoS}}[n]\right|^{2}$. Thus, problem (17) can be equivalently converted into the following optimization problem:
\begin{align*}
	&\max _{\mathbf{\Psi}} \left|\sum_{m=1}^{M} h_{m}^{\mathrm{LoS}}[n]+h_{D}^{\mathrm{LoS}}[n]\right|^{2} \tag{12}\\
	&\quad\text{s.t. (10d).} 
\end{align*}}
	\textbf{Proposition 1:} The optimal phase shift for the $m$th element of the IOS is
	\begin{equation}
		\psi_{m}[n]=\frac{2 \pi}{\lambda}\left(d_{U A V,G}[n]-d_{U A V, m}[n]-d_{m,G}\right). \tag{13}
	\end{equation}
	\emph{Proof:} See Appendix A. $\hfill\blacksquare$
	\subsection{UAV Trajectory Optimization}
	With the optimal phase shift obtained by (13), the UAV trajectory optimization problem can be expressed as
	\begin{small}
\begin{align*}
	\max _{\mathbf{Q}} &\frac{1}{N} \sum_{n=1}^{N} \log _{2}\left(1+ \eta\zeta^{2}[n]\left|\sqrt{\frac{\kappa}{1+\kappa}} e^{ \dfrac{-jd_{U,G}[n]}{\lambda}}+\sqrt{\frac{1}{1+\kappa}} h_{S S}\right|^{2} \right)  \tag{14}\\
	&\text { s.t. } \text{(10b)-(10c)},
\end{align*}
\end{small}where $\zeta[n]=\sum_{m=1}^{M} \frac{J_m\beta_{m}\left|x[n]-x_{m}\right|^{3}}{d_{{UAV}, m}^{4}[n]} +\frac{K}{d_{{UAV},G}^{\alpha / 2}[n]}$, $J_m=\frac{\lambda \sqrt{G^{\mathrm{tx}} G^{\mathrm{rx}} G_{m} \delta_{z} \delta_{y}\left|\gamma_{m}\right|^{2}}}{(4 \pi)^{\frac{3}{2}}}$, $K=\sqrt{G^{tx}G^{rx}}$ and $\beta_m=\frac{K^D_m}{d_{m,G}^{}}$. It is observed that problem (14) is still non-convex with respect to $\mathbf{Q}$. To solve this problem, we have the following proposition to convert this problem into an equivalent problem.
	
\textbf{Proposition 2.} Problem (14) is equivalent to the following problem: 
	\begin{align*}
		\max _{\mathbf{Q},\mathbf{s},\mathbf{u},\mathbf{v}}& \frac{1}{N} \sum_{n=1}^{N} \log _{2}\left(1+\eta\left(\sum_{m=1}^{M} J_m \beta_{m} \frac{e^{s \mid n]}}{e^{u_m[n] }}+K e^{v[n]}\right)^{2}\right) \tag{15a} \\
		\text { s.t. }&e^{s[n]} \leq|x[n]|^{3}, \tag{15b} \\
		&e^{u_m[n]} \geq d_{U , m}^{4}[n],  \tag{15c} \\
		&e^{v[n]} \leq d_{U ,G}^{-\alpha / 2}[n], \tag{15d}\\
		&\text{(10b)-(10c)}. 
	\end{align*}
where $\mathbf{s}=\{s[n]\}_{n=1}^{N}$, $\mathbf{u}={\{u_m[n]\}_{n=1}^{N}}_{m=1}^M$, and $\mathbf{v}=\{v[n]\}_{n=1}^{N}$.
\emph{Proof:} See Appendix B. $\hfill\blacksquare$
	
	Note that, after the variable replacement,  the objective function (15a) is transformed to a log-sum-exp function which is  convex  \cite{2004Convex}.   However, constraints (15b)-(15d) are still non-convex. Since the first-order Taylor approximation of a convex function is a global underestimator, it can be applied at any local points  $s^{(l)}[n]$, $u_{m}^{(l)}[n]$, $v^{(l)}[n]$ and $|x^{(l)}[n]|$ in the $l$th iteration for (15a)-(15d) i.e., 
	\begin{small}\begin{align*}
		\log _{2}\left(1+\eta\left(\sum_{m=1}^{M} J_m \beta_{m} \frac{e^{s[n]}}{e^{u_m[n]}}+Ke^{v[n]}\right)^{2}\right) &\geq \log _{2} A^{(l)}[n]+\frac{B^{(l)}[n]}{A^{(l)}[n] \ln 2}\left(s[n]-s^{(l)}[n]\right)
		+\sum_{m=1}^{M}\frac{C_{m}^{(l)}[n]}{A^{(l)}[n] \ln 2}\left(u_m[n]-u_{m}^{(l)}[n]\right)\\
		&\quad+\frac{D^{(l)}[n]}{A^{(l)}[n] \ln 2}\left(v[n]-v^{(l)}[n]\right),\tag{16} \\
		&\left|x^{(l)}[n]\right|^{3}+\frac{3 x^{(l)}[n]^{3}}{\left|x^{(l)}[n]\right|}\left(x[n]-x^{(l)}[n]\right)\ge e^{s[n]}, \tag{17}\\
		&e^{\frac{u_{m}^{(l)}[n]}{2}}+\frac{e^{\frac{u_{m}^{(l)}[n]}{2}}}{2}\left(u_m[n]-u_{m}^{(l)}[n]\right) \ge d^2_{U,m}[n], \tag{18}\\
		&e^{\frac{-4 v^{(l)}[n]}{\alpha}}-\frac{4 e^{\frac{-4 v^{(l)}[n]}{\alpha}}}{\alpha}\left(v[n]-v^{(l)}[n]\right) \ge d^2_{U,G}[n], \tag{19}\\\end{align*}\end{small}$\text{where}$\\
	 $A^{(l)}[n]=1+\eta\left(\left(\sum_{m=1}^{M} J_m \beta_{m} {\frac{e^{s^{(l)}[n]}}{e^{u_{m}^{(l)}[n]}}}\right)^{2}+K^{2} e^{2 v^{(l)}[n]}\right.\left.+2 K \sum_{m=1}^{M}J_m \beta_{m} {\frac{e^{s^{(l)}[n]+v^{(l)}[n]}}{e^{u_{m}^{(l)}[n]}}}\right)$, \\
	$B^{(l)}[n]=\eta\left(2\left(\sum_{m=1}^{M} J_m\beta_{m} \frac{e^{s^{(l)}[n]}}{e^{u_{m}^{(l)}[n]}}\right)^{2}+2 K \sum_{m=1}^{M} J_m\beta_{m} \frac{e^{s^{(l)}[n]+v^{l}[n]}}{e^{u_{m}^{(l)}[n]}}\right)$,\\
$C_{m}^{(l)}[n]=\eta\left(-2\left(\sum_{m=1}^{M} J_m \beta_{m}\frac{e^{s^{(l)}[n]}}{e^{u_{m}^{(l)}[n]}}\right)^{}J_m\beta_m\dfrac{e^{s^{(l)}[n]}}{e^{u_m^{(l)}[n]}} \right.
 \left.-2J_mK  \beta_{m} \frac{e^{s^{(l)}[n]+v^{(l)}[n]}}{e^{u_{m}^{(l)}[n]}}\right)$,\\
$ D^{(l)}[n]=\eta\left(2 K^{2} e^{2 v^{(l)}[n]}+2 K \sum_{m=1}^{M}J_m \beta_{m} \frac{e^{s^{(l)}[n]+v^{(l)}[n]}}{e^{u_{m}^{(l)}[n]}}\right).$
	With (16)-(19), problem (15) can be approximated as
	\begin{align*}
		\max_{\mathbf{Q},\mathbf{s},\mathbf{u},\mathbf{v}}\quad&\dfrac{1}{N}\sum_{n=1}^{N}\frac{B^{(l)}[n]}{A^{(l)}[n] \ln 2}s[n]+\sum_{m=1}^{M}\frac{C_{m}^{(l)}[n]}{A^{(l)}[n] \ln 2}u_m[n]+\frac{D^{(l)}[n]}{A^{(l)}[n] \ln 2}v[n], \tag{20} \\
		\text { s.t. }&\text{(10a)-(10b), (17)-(19)}. 
	\end{align*}
	Problem (20) is now a convex optimization problem and can therefore be efficiently solved by a standard CVX solver \cite{jj}.
%		\begin{algorithm}[t]
%		\caption{Proposed algorithm for problem (10)} %算法的名字11
%		\begin{algorithmic}[1]
%			\State $\mathbf{Initialization:}$Initialize  $\mathbf{Q}^{(0)}$, $\mathbf{\Psi}^{(0)}$, $\mathbf{s}^{(0)}$, $\mathbf{u}^{(0)}_m$, $\mathbf{x}^{(0)}$, $\mathbf{v}^{(0)}$ and
%			iteration number $l = 0$. Set $\bar{R}$ with (9) and given
%			$(\mathbf{Q}^{(0)}, \mathbf{\Psi}^{(0)})$.
%			\Repeat
%			\State Set $l \leftarrow  l+1$;
%			\State Update $(\mathbf{Q}^{(l)}, \mathbf{s}^{(l)}, \mathbf{u}_m^{(l)}, \mathbf{x}^{(l)}, \mathbf{v}^{(l)})$ by solving problem (19);
%			\State With given $\mathbf{Q}^{(l)}$, update $\mathbf{\Psi}^{(l)}$ by employing (12);
%			\State With given $(\mathbf{Q}^{(l)},\mathbf{\Psi}^{(l)})$, update $\bar{R}^{(l)}$ by employing (9)
%			\Until{$\frac{\bar{R}^{{(l)}}-\bar{R}^{{(l)}}}{\bar{R}^{{(l)}}}<\mu$}.
%		\end{algorithmic}
%	\end{algorithm}
	\subsection{Overall Algorithm}
	By using our proposed algorithm, problem (10) can be efficiently solved by alternately optimizing variables $\mathbf{\Psi}$ and $\mathbf{Q}$, while its solution converges to a preset accuracy $\mu$. \textcolor{blue}{Since the two subproblems are solved by applying CVX via the standard interior point method, their computational complexity can be obtained as $O\left((M N)^{3.5} \log (1 / \mu)\right)$ and $O\left((4 N+M N)^{3.5} \log (1 / \mu)\right)$, respectively. Besides, the computational complexity of the alternating optimization is $O(\log (1 / \mu))$. Thus, the total computational complexity of our proposed algorithm is in the order of $O\left((4 N+M N)^{3.5} \log ^{2}(1 / \mu)\right)$.}
%		\begin{figure}
%		\centering
%		\includegraphics[width=80mm]{lyf1.png}\\
%		\caption{UAV trajectories by different elements. }
%		\label{fig:env}
%	\end{figure} 
%	
%	
%	\begin{figure}
%		\centering
%		\includegraphics[width=80mm]{lyf2.png}\\
%		\caption{Average rate by different elements verus
%			T. }
%		\label{fig:env}
%	\end{figure} 

	\section{Simulation Results}
	In this section, we evaluate the performance of the IOS-assisted UAV-enabled communications based on the proposed algorithm (denoted by IA scheme) and compare it  with the following benchmark solutions: 1) RIS-assisted UAV communications (denoted by RA scheme) proposed in \cite{8959174}, where the IOS in our proposed system is replaced by an RIS; 2) IOS-assisted  UAV communications with the fixed trajectory (denoted by IA-FT scheme), where the UAV takes off from $\mathbf{q}_0$ to G at $v_{max}$, then hovers above G as long as possible, and finally flies towards $\mathbf{q}_F$ at $v_{max}$ by the end of $T$; 3) Conventional UAV communication (denoted by CUC scheme) without the aid of IOS or RIS.  We assume that the IOS and G are located at (0, 0, 40) m and (-100, -20, 0) m, respectively. Other simulation parameters are set as: $\mathbf{q}_0 = [-400,20]$ m, $\mathbf{q}_F = [400,20]$ m, $v_{max} = 25 $ m/s, $\mu=0.0001$, $G_{tx}=G_{rx}=1$, $\delta_t$ = 1 s, $\kappa=3$, $P = 0.1$ W, $\alpha=5$, $\sigma^2=-80$ dBm, $\lambda=0.05$ m, $\gamma_{m}=1$, $\epsilon=3.55$ \cite{2013}.

	Fig. 1(a) presents the different trajectories of the UAV by various algorithms with $T=150$ s and $M=6000$.   Comparing to the RA scheme that the UAV only hovers at the location on the left side of the RIS, the UAV in the IA scheme hovers close to both sides of the IOS to gain the benefit of omni-directional rate enhancement. The reason behind this observation is that the IOS is not only able to reflect the arrived signal from the UAV like the RIS, but also to transmit the signal. This results in higher rate compared to the RA scheme, which can be verified in the following Fig. 1(b). 
	\begin{figure}[h]
		\centering
		\subfigure[UAV trajectories]{
			\begin{minipage}[t]{0.48\linewidth}
				\centering
				\includegraphics[width=3in]{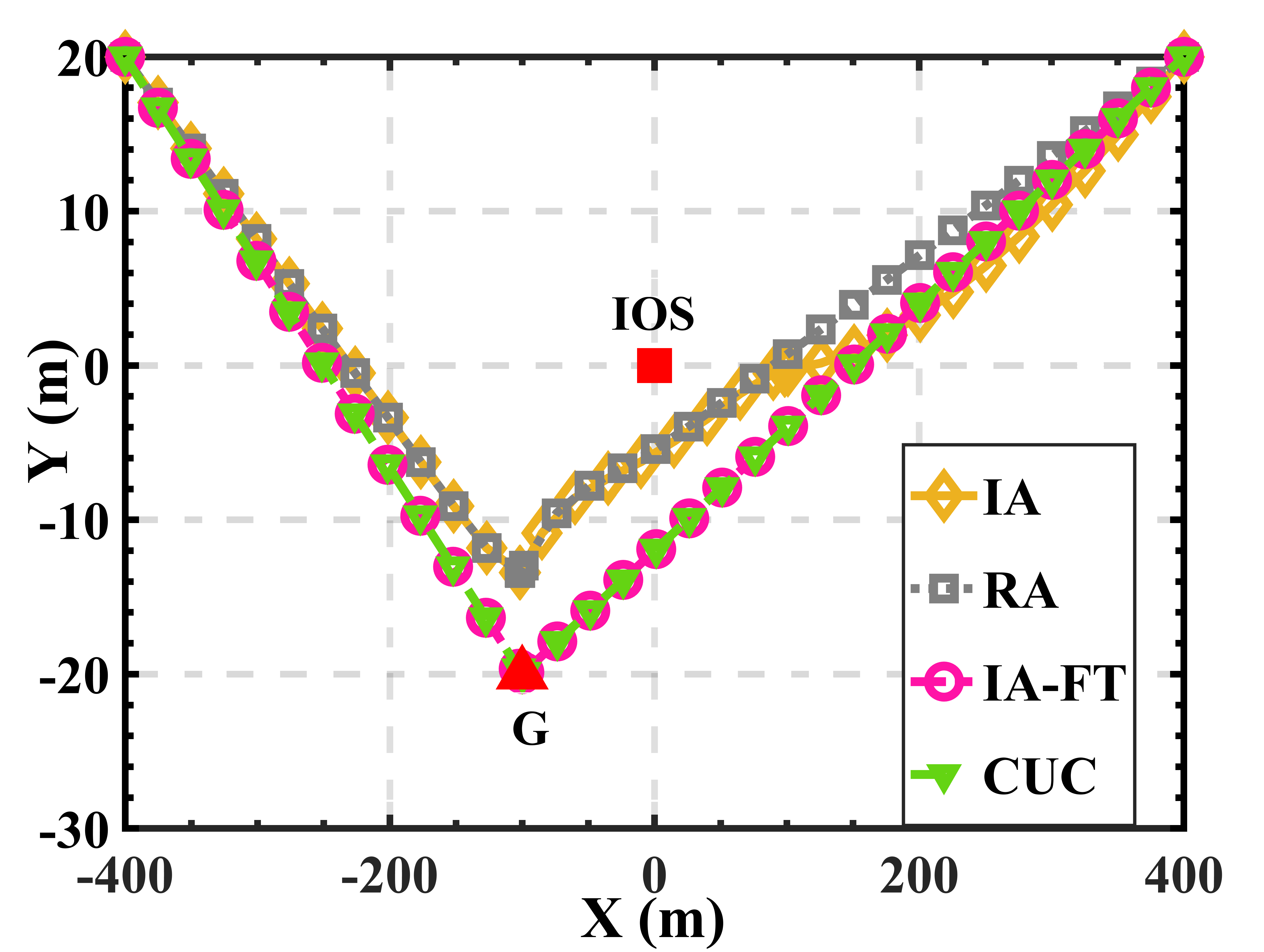}
				%\caption{fig1}
			\end{minipage}%
		}%
		\subfigure[Average rates versus $T$]{
			\begin{minipage}[t]{0.5\linewidth}
				\centering
				\includegraphics[width=3in]{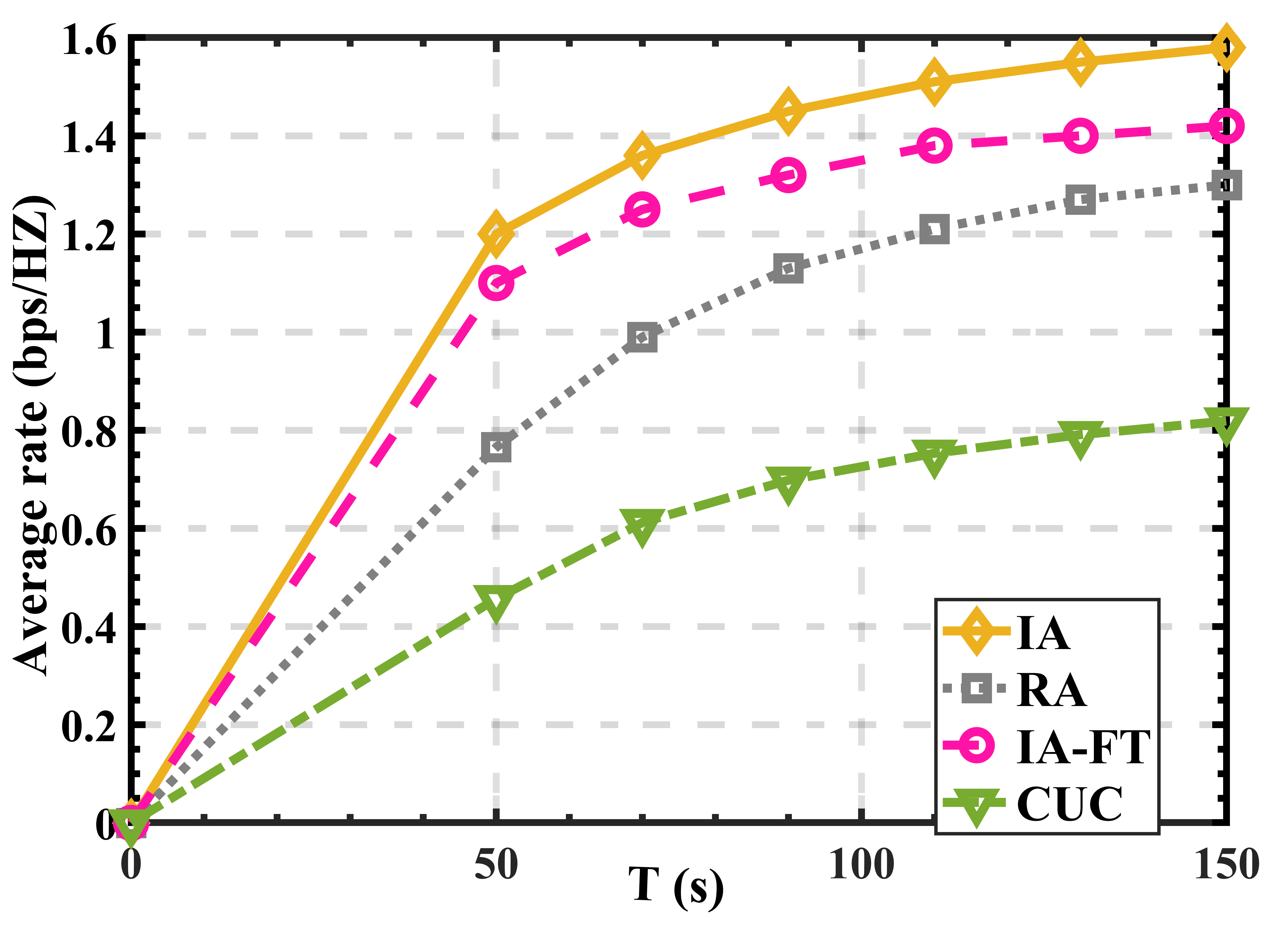}
				%\caption{fig2}
			\end{minipage}%
		}%
		\centering
		\caption{UAV Trajectories and average achievable rates by different schemes.} 
	\end{figure}
%	 Furthermore, the UAV trajectory  in the CUC scheme is identical to that in the IA-FT  scheme, since the ground node G can obtain the higher average achieve rate along this optimized trajectory.
%	 For cellular communications, UAV hovers at the same position as the fixed trajectory algorithm, i.e., directly above the user. In conventional cellular communication, the faster the UAV approaches the hovering position directly above the user, the higher the average rate. Therefore,Furthermore, we can observe that in the RIS-assisted scheme, the UAV flies to the endpoint at maximum speed after reaching the right side of the IOS. In contrast, in the IOS-assisted scheme, the UAV hovers long enough on the right side and then flies towards the endpoint.
%		\begin{figure}
%	\centering
%	\includegraphics[width=80mm]{mmm.png}\\
%	\caption{UAV trajectories by different elements. }
%	\label{fig:env}
%\end{figure} 
%
%
%\begin{figure}
%	\centering
%	\includegraphics[width=80mm]{D.png}\\
%	\caption{Average rate by different elements verus
%		T. }
%	\label{fig:env}
%\end{figure}  
	Fig. 1(b) illustrates the average achievable rates by different schemes with $M=6000$ versus $T$. It is observed that the average achievable rates are significantly improved with the increase of  $T$, since the UAV can fly to more favorable hovering locations to  transmit more information. Particularly, the RA scheme achieves a significant rate gain over the CUC scheme, which confirms that the proper deployment of the RIS can indeed enhance the achievable rate of the system effectively.  Our proposed IA scheme obtains more significant rate improvement compared to the RA scheme. This indicates that properly designing the UAV trajectory and the IOS's phase shift can benefit from the flexible mobility of the UAV as well as the full-dimensional rate enhancement with the assistance of the IOS, thus unlocking the full potential of IOS-aided UAV communications. 
	\begin{figure}[h]
		\centering
		
		\subfigure[UAV trajectories]{
			\begin{minipage}[t]{0.48\linewidth}
				\centering
				\includegraphics[width=3in]{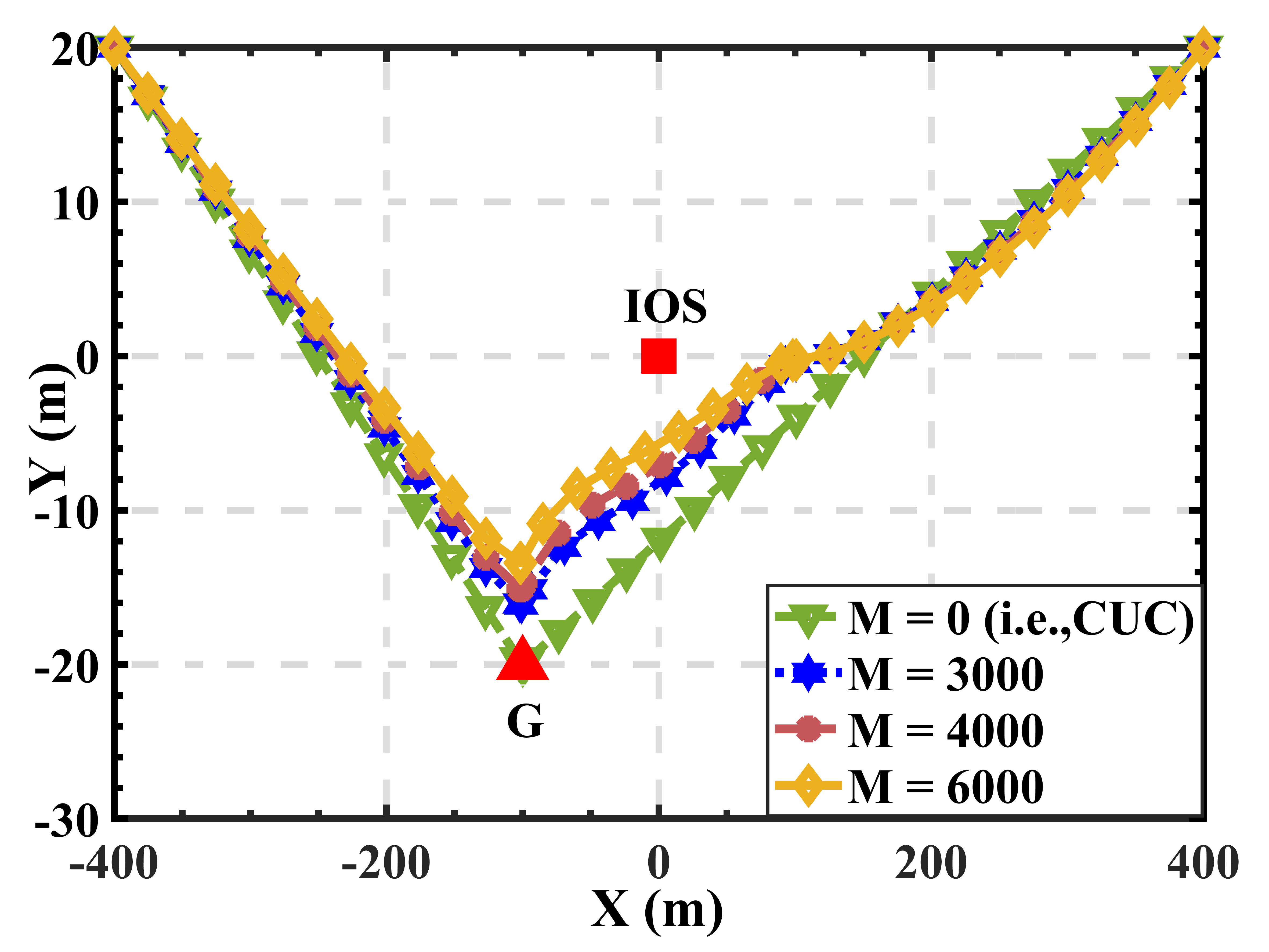}
				%\caption{fig1}
			\end{minipage}%
		}%
		\subfigure[Average rates versus $T$]{
			\begin{minipage}[t]{0.5\linewidth}
				\centering
				\includegraphics[width=3in]{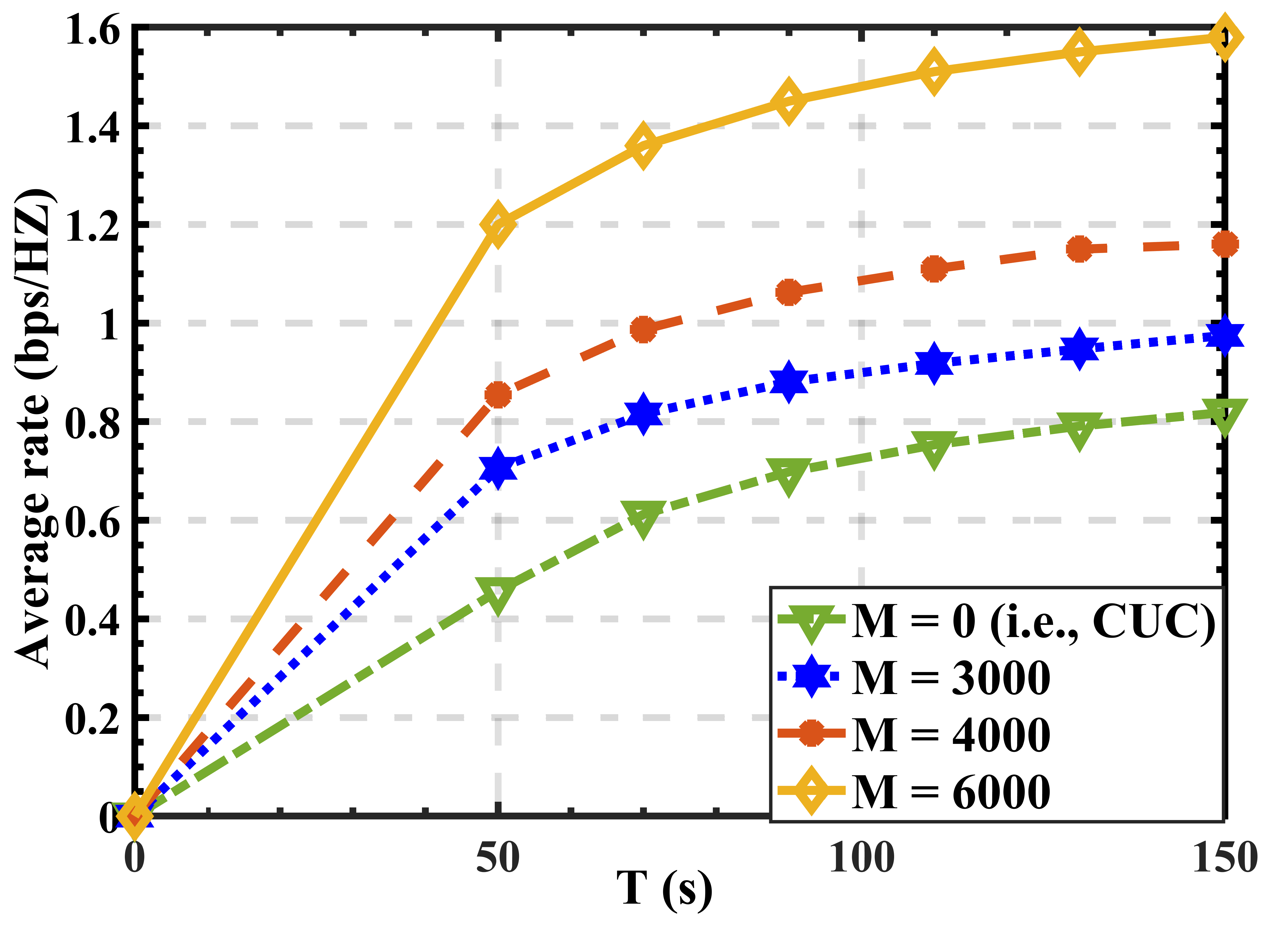}
				%\caption{fig2}
			\end{minipage}%
		}%
		\centering
		\caption{UAV Trajectories and average achievable rates by different $M$.} 
	\end{figure}
	Fig. 2(a) shows the different UAV trajectories by our proposed IA scheme versus $M$ for $T$ = 150 s. Note that with the increase of $M$, the UAV's trajectory and its hovering locations tend to approach the IOS due to the greater benefit that the IOS can provide from its both sides, which can be verified in Fig. 2(b). Fig. 2(b) shows the average achievable rates for different $M$ versus $T$. As expected, the rate performance is improved significantly when equipping with more elements in the IOS due to the larger passive beamforming gain.	
	\section{Conclusion}
	In this paper, we investigated the potential for increasing the rate performance by leveraging the novel reflective-transmissive IOS. The objective is to maximize the average achievable rate.  We proposed an efficient algorithm to obtain a suboptimal solution by alternately optimizing the IOS's phase shift and the UAV trajectory. Simulation results have shown that the IOS-aided scheme achieves noteworthy gain as compared to the conventional UAV-enabled communications with/without the aid of a traditional RIS. This demonstrates the potential for using the IOS to provide the full coverage of
	communication services in future 6G networks.
%	 Moreover, using the IOS for UAV communication
%	systems also offers many advantages compared with using the RIS, such as the UAV is not restricted to flying in
%	front of the RIS and can have a more flexible trajectory and hover point design.
	\appendices
	\section{Proof of Proposition 1}
	\textcolor{blue}{To maximize $|h[n]|$ shown in problem (12), the triangle inequality can be applied, i.e.,
	\begin{small}
	\begin{align*}
		&\left|\sum_{m=1}^{M}h_{m}^{\rm{LoS}}[n]+h_{D}^{\operatorname{L o S}}[n]\right|\overset{a}{\le}\sum_{m=1}^{M}\left|h_{m}^{\rm{LoS}}[n]\right|+\left|h_{D}^{\rm {L o S}}[n]\right|=\sum_{m=1}^{M}\bigg|\\
		& \left.\frac{J_m \beta_{m}\left|x[n]-x_{m}\right|^{3}}{d_{{U}, m}^{4}[n]}\exp \left(\frac{-j 2 \pi\left(d_{U, m}[n]+d_{m, G}[n]\right)}{\lambda}-j\psi_{m}[n]\right)\right| \\
		&+\left| \frac{K}{d_{{U}, G}^{\alpha / 2}[n]}\exp \left(\frac{-j 2 \pi d_{U, G}[n]}{\lambda}\right)\right| ,\tag{21}
	\end{align*}
\end{small}where $J_m=\frac{\lambda \sqrt{G^{\mathrm{tx}} G^{\mathrm{rx}} G_{m} \delta_{z} \delta_{y}\left|\gamma_{m}\right|^{2}}}{(4 \pi)^{\frac{3}{2}}}$, $K=\sqrt{G^{tx}G^{rx}}$ and $\beta_m=\frac{K^D_m}{d_{m,G}^{}}$. Note that (a) holds with equality if and only if  
	$\psi_{m}[n]+\frac{2 \pi\left(d_{U, m}[n]+d_{m,G}\right)}{\lambda}=\frac{2 \pi\left(d_{U,G}[n]\right)}{\lambda}$ \cite{9400768}. Therefore, the
	optimal solution to problem (12) is
	\begin{equation}
		\psi_{m}[n]=\frac{2 \pi}{\lambda}\left(d_{U,\textsc{G}}[n]-d_{U, m}[n]-d_{m,G}\right).\tag{22}
	\end{equation}}

	% you can choose not to have a title for an appendix
	% if you want by leaving the argument blank
	\section{Proof of Proposition 2}
	To solve problem (14), we first apply Euler's formula to calculate $|h[n]|^{2}$ as below
	\begin{small}
		\begin{align*}
		|h[n]|^{2} &=\zeta[n]^{2}\left|\sqrt{\frac{\kappa}{1+\kappa}} \exp (\dfrac{-j d_{U,G}[n]}{\lambda})+\sqrt{\frac{1}{1+\kappa}} h_{S S}\right|^{2} \\
		&=\zeta[n]^{2}\left(\left|\sqrt{\frac{\kappa}{1+\kappa}} \cos (\dfrac{-j d_{U,G}[n]}{\lambda})+\sqrt{\frac{1}{1+\kappa}} \operatorname{Re}\left(h_{S S}\right)\right|^{2}\right.\\
		&\quad\left.+|j|^{2}\left|\sqrt{\frac{\kappa}{1+\kappa}} \sin (\dfrac{-j d_{U,G}[n]}{\lambda})+\sqrt{\frac{1}{1+\kappa}} \operatorname{Im}\left(h_{S S}\right)\right|^{2}\right) \\
%		&=\frac{\zeta[n]^{2}}{1+\kappa}\left(\kappa \cos ^{2}(\dfrac{ d_{U,G}[n]}{\lambda})+\kappa \sin ^{2} (\dfrac{d_{U,G}[n]}{\lambda})+\operatorname{Im}\left(h_{S S}\right)^{2}\right.\\
%		&\quad+\left.2\sqrt{\kappa}\operatorname{Re}\left(h_{S S}\right) \cos(\dfrac{ d_{U,G}[n]}{\lambda})-2 \sqrt{\kappa} \operatorname{Im}\left(h_{S S}\right) \sin(\dfrac{ d_{U,G}[n]}{\lambda})\right. \\
%		&\quad\left.+\operatorname{Re}\left(h_{S S}\right)^{2}\right)\\
		&=\frac{\zeta[n]^{2}}{1+\kappa}\left(\kappa+\left|h_{S S}\right|^{2}+2 \sqrt{\kappa} \operatorname{Re}\left(h_{S S}\right) \cos (\dfrac{ d_{U,G}[n]}{\lambda})\right.\\
		&\quad\left.-2 \sqrt{\kappa} \operatorname{Im}\left(h_{S S}\right) \sin(\dfrac{ d_{U,G}[n]}{\lambda})\right), \tag{23}
	\end{align*}
\end{small}where $\zeta[n]=\sum_{m=1}^{M} \frac{J_m\left|x[n]-x_{m}\right|^{3}}{d_{{U}, m}^{4}[n]} \frac{\left|x_{G}-x_{m}\right|^{3}}{d_{m, G}^{4}}+\frac{K}{d_{{U},G}^{\alpha / 2}[n]}$. Therefore, the channel power gain from the UAV to G via IOS can be expressed as
\begin{small}
	\begin{align*}
		\mathbb{E}\left(|h[n]|^{2}\right)=\sum_{m=1}^{M} \frac{J_m\left|x[n]-x_{m}\right|^{3}}{d_{\mathrm{UAV}, m}^{4}[n]} \frac{\left|x_{G}-x_{m}\right|^{3}}{d_{m, G}^{4}}+\frac{K}{d_{{UAV},G}^{\alpha / 2}[n]}, \tag{24}
	\end{align*}
\end{small}
where $\mathbb{E}(\cdot)$ is the expectation operator.
	With (24), problem (10) can be transformed  to
	\begin{small}\begin{align*}
		\max _{\mathbf{Q}} &\frac{1}{N} \sum_{n=1}^{N} \log _{2}\left(1+  \eta \left(\sum_{m=1}^{M} \frac{J_m \beta_m\left|x[n]\right|^{3}}{d_{{U}, m}^{4}[n]} +\frac{K}{d_{{U}, G}^{\alpha / 2}[n]}\right)^{2}\right)  \tag{25}\\
		&\text { s.t. } \text{(10b)-(10c)}. 
	\end{align*}\end{small}
	To solve this non-convex problem, we further introduce the variables $ e^{s[n]}$, $ e^{u_m[n]} $ and $e^{v[n]}$ to replace $|x[n]|^3$, $d_{{U}, m}^{4}[n]$ and $d_{{U}, G}^{4}[n]$, respectively. Then, we have problem (15) which completes the proof.

	% use section* for acknowledgment

	% Can use something like this to put references on a page
	% by themselves when using endfloat and the captionsoff option.
	\ifCLASSOPTIONcaptionsoff
	\newpage
	\fi

	% trigger a \newpage just before the given reference
	% number - used to balance the columns on the last page
	% adjust value as needed - may need to be readjusted if
	% the document is modified later
	%\IEEEtriggeratref{8}
	% The "triggered" command can be changed if desired:
	%\IEEEtriggercmd{\enlargethispage{-5in}}
	
	% references section
	
	% can use a bibliography generated by BibTeX as a .bbl file
	% BibTeX documentation can be easily obtained at:
	% http://mirror.ctan.org/biblio/bibtex/contrib/doc/
	% The IEEEtran BibTeX style support page is at:
	% http://www.michaelshell.org/tex/ieeetran/bibtex/
	%\bibliographystyle{IEEEtran}
	% argument is your BibTeX string definitions and bibliography database(s)
	%\bibliography{IEEEabrv,../bib/paper}
	%
	% <OR> manually copy in the resultant .bbl file
	% set second argument of \begin to the number of references
	% (used to reserve space for the reference number labels box)
	%\textit{\begin{thebibliography}{1}
	%
	%\bibitem{IEEEhowto:kopka}
	%H.~Kopka and P.~W. Daly, \emph{A Guide to \LaTeX}, 3rd~ed.\hskip 1em plus
	%  0.5em minus 0.4em\relax Harlow, England: Addison-Wesley, 1999.
	%  S. Chen, Y. -C. Liang, S. Sun, S. Kang, W. Cheng and M. Peng, "Vision, Requirements, and Technology Trend of 6G: How to Tackle the Challenges of System Coverage, Capacity, User Data-Rate and Movement Speed," in IEEE Wireless Communications, vol. 27, no. 2, pp. 218-228, April 2020, doi: 10.1109/MWC.001.1900333.
	%
	%\end{thebibliography}
	%}
	\bibliographystyle{IEEEtran}
	\bibliography{mycite}
	% biography section
	% 
	% If you have an EPS/PDF photo (graphicx package needed) extra braces are
	% needed around the contents of the optional argument to biography to prevent
	% the LaTeX parser from getting confused when it sees the complicated
	% \includegraphics command within an optional argument. (You could create
	% your own custom macro containing the \includegraphics command to make things
	% simpler here.)
	%\begin{IEEEbiography}[{\includegraphics[width=1in,height=1.25in,clip,keepaspectratio]{mshell}}]{Michael Shell}
	% or if you just want to reserve a space for a photo:
	%
%	\begin{IEEEbiography}{Bin Duo}
%		
%	\end{IEEEbiography}
%	\begin{IEEEbiography}{Yifan Liu}
%	
%\end{IEEEbiography}
%	\begin{IEEEbiography}{Qingqing Wu}
%	
%\end{IEEEbiography}
%	\begin{IEEEbiography}{Xiaojun Yuan}
%	
%\end{IEEEbiography}
	%
	%% if you will not have a photo at all:
	%\begin{IEEEbiographynophoto}{John Doe}
	%Biography text here.
	%\end{IEEEbiographynophoto}
	%
	%% insert where needed to balance the two columns on the last page with
	%% biographies
	%%\newpage
	%
	%\begin{IEEEbiographynophoto}{Jane Doe}
	%Biography text here.
	%\end{IEEEbiographynophoto}
	
	% You can push biographies down or up by placing
	% a \vfill before or after them. The appropriate
	% use of \vfill depends on what kind of text is
	% on the last page and whether or not the columns
	% are being equalized.
	
	%\vfill
	
	% Can be used to pull up biographies so that the bottom of the last one
	% is flush with the other column.
	%\enlargethispage{-5in}

	% that's all folks
\end{document}